\pdfoutput=1
\documentclass[10pt, 
a4paper, 
oneside, 
headinclude,footinclude, 
BCOR5mm, 
]{article}

\usepackage[
nochapters, 
beramono, 
eulermath,
pdfspacing, 
dottedtoc 
]{classicthesis} 

\usepackage[T1]{fontenc} 

\usepackage[utf8]{inputenc} 

\usepackage{graphicx} 
\graphicspath{{Figures/}} 

\usepackage{enumitem} 

\usepackage{lipsum} 

\usepackage{subfig} 

\usepackage{amsmath,amssymb,amsthm} 

\usepackage{varioref} 

\usepackage[left=2cm,right=2cm,top=2cm,bottom=2cm]{geometry} 

\usepackage{algorithm}
\usepackage{algorithmic} 

\usepackage{url} 

\theoremstyle{definition} 

\theoremstyle{plain} 

\theoremstyle{remark} 

\hypersetup{
draft, 
colorlinks=true, breaklinks=true, bookmarks=true,bookmarksnumbered,
urlcolor=webbrown, linkcolor=RoyalBlue, citecolor=webgreen, 
pdftitle={}, 
pdfauthor={\textcopyright}, 
pdfsubject={}, 
pdfkeywords={}, 
pdfcreator={pdfLaTeX}, 
pdfproducer={LaTeX with hyperref and ClassicThesis} 
} 

\title{\normalfont\spacedallcaps{A parallel implementation of the Synchronised Louvain Method}} 

\author{Benjamin Chiêm$^{1}$ \& Andine Havelange$^{1}$ \& Paul Van Dooren$^{2}$} 

\date{\today} 

\begin{document}


\renewcommand{\sectionmark}[1]{\markright{\spacedlowsmallcaps{#1}}} 
\lehead{\mbox{\llap{\small\thepage\kern1em\color{halfgray} \vline}\color{halfgray}\hspace{0.5em}\rightmark\hfil}} 

\pagestyle{scrheadings} 

\maketitle 





\section*{Abstract} 

Community detection in networks is a very actual and important field of research with applications in many areas. But, given that the amount of processed data increases more and more, existing algorithms need to be adapted for very large graphs. \\
The objective of this project was to parallelise the Synchronised Louvain Method, a community detection algorithm developed by A. Browet, in order to improve its performances in terms of computation time and thus be able to faster detect communities in very large graphs. \\
To reach this goal, we used the API OpenMP to parallelise the algorithm and then carried out performance tests. \\
We studied the computation time and speedup of the parallelised algorithm and were able to bring out some qualitative trends. We obtained a great speedup, compared with the theoretical prediction of Amdahl's law. \\
To conclude, using the parallel implementation of Browet's algorithm on large graphs seems to give good results, both in terms of computation time and speedup. Further tests should be carried out in order to obtain more quantitative results.


\section*{Acknowledgement}
We would like to sincerely thank Etienne Huens for his technical help and his pieces of advice about all coding aspects, and Arnaud Browet for his explanations about his innovative algorithm.


{\let\thefootnote\relax\footnotetext{\textsuperscript{1} \textit{Ecole Polytechnique de Louvain, Université Catholique de Louvain, Louvain-la-Neuve, Belgium}}}

{\let\thefootnote\relax\footnotetext{\textsuperscript{2} \textit{(Supervisor) Department of Mathematical Engineering, Ecole Polytechnique de Louvain, Université Catholique de Louvain, Louvain-la-Neuve, Belgium}}}


\newpage
\setcounter{tocdepth}{2} 

\tableofcontents 

\listoffigures 

\listoftables 


\newpage 


\section{Introduction}

Since the last decade, the study of large networks has known a big boost of interest. Graph theory provides interesting new insights for many studies. This is why we need new algorithms to analyse the structure of these networks.\\

In particular, an interesting question to be raised is: does a certain network contain some kind of communities? This is precisely the question that the Synchronised Louvain method aims to answer. Developed by Arnaud Browet, this algorithm is based on the well-known Louvain method. The goal of our project is to implement a parallel version of Browet's algorithm in order to improve its performances in terms of computation time and therefore, to be able to quickly detect communities in very large networks.\\

\begin{figure}[h!]
    \centering   
    \includegraphics[scale=0.8]{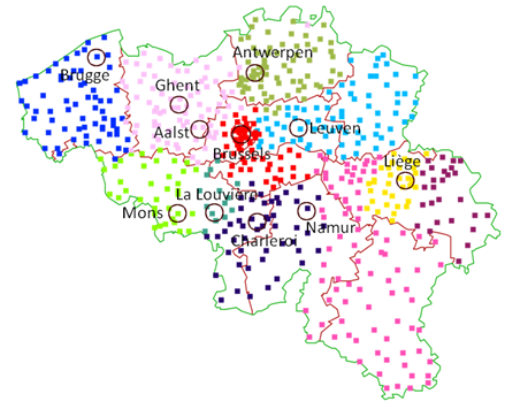}
    \caption[Communication communities in Belgium]{Communication communities in Belgium. (Source: Guigourès, R., Boullé, M., Rossi, F. (s.d.). \textit{Segmentation géographique par étude d’un journal d’appels téléphoniques}. \url{http://apiacoa.org/publications/2011/guigouresboulleetal2011segmentation-geographique.pdf})}
    \label{CommBelgium}
\end{figure}

Accelerating community detection algorithms could be of great interest in the domain of telecommunications (Figure \ref{CommBelgium}) for example, but also in others. In fact, many new studies are trying to apply such tools provided by graph theory to various domains such as neuroscience, social or even gene regulatory networks. The amount of data to process in these cases is quite huge; it is thus well justified to search for the fastest algorithm.\\

In this document, we will briefly explain some general notions about communities and the different steps of the Synchronised Louvain method, then we will present what is parallel computing. After that, we will detail the way we chose to implement our parallel version and what tools we used. Finally, we will develop our results and give some explanations about them.

\section{General notions}

\subsection{Definition of communities}

There is no unique definition of communities. They can informally be defined as sets of nodes with a high internal density (which is defined as the ratio between the actual number of edges in this set and the maximal number of possible edges) and a low external density with the rest of the nodes not being in the community. In other words, a community is a set of nodes who are strongly interconnected and weakly connected with the rest of the network. A typical example is the study of the belgian communication network (see Figure \ref{CommBelgium}) in which communities can be detected, some of them corresponding to provinces of Belgium. \\

This definition is unfortunately too informal to allow us to compute what would be an optimal partition of nodes into communities in a given graph. Most of the time, people use functions that measure the quality of a partition of the nodes into communities, called fitness measures or quality functions. The goal is then to find a community structure maximising the quality function.\\

There exist various quality functions and some of them are based on the concept of energy of a partition. It has been developed by Reichardt and Bornholdt and will be presented in the following subsection.

\subsection{Energy of a partition}
Reichardt and Bornholdt interpret the community detection problem as finding the ground state of a spin glass, that is a disordered magnet. A spin state $\sigma_i$ is assigned to each node of the graph. This spin state represents the community assignment of the node. So, if the nodes are partitioned into $c$ communities, $\sigma_i \in \{1,2,...,c\}$.  \\

In an optimal partition of nodes into communities, edges should ideally connect vertices in the same spin state, i.e. belonging to the same community. And there should be as little as possible edges linking nodes in different spin states. This observation leads to the definition of an energy function rewarding existing edges between vertices in the same spin state and penalising other existing edges. \\

If the adjacency matrix of the graph is denoted by $A$, for any existing edge, $A(i,j) \neq 0$, the partition is 
\renewcommand\labelitemi{\textbullet}
\begin{itemize}
\item  rewarded by $a_{ij} > 0$ if the nodes $i$ and $j$ are in the same spin state, i.e. $\sigma_i = \sigma_j$
\item penalised by $c_{ij} > 0$ if the nodes are in different spin states $\sigma_i \neq \sigma_j$
\end{itemize}
Non-existing edges must also be taken into account in a similar way. More precisely, for any non-existing edge in the graph, $A(i,j)=0$, the partition is 
\begin{itemize}
\item penalised by $b_{ij} > 0$ if the nodes $i$ and $j$ are in the same spin state, i.e. $\sigma_i = \sigma_j$
\item rewarded by $d_{ij} > 0$ if the nodes are in different spin states $\sigma_i \neq \sigma_j$
\end{itemize}

Denoting the Kronecker delta by $\delta(\sigma_i,\sigma_j)$, i.e.
\begin{eqnarray*}
  \delta(\sigma_i,\sigma_j) &=&
  \begin{cases}
    1 & \text{if } \sigma_i = \sigma_j \\
    0 & \text{if } \sigma_i \neq \sigma_j
  \end{cases}
\end{eqnarray*}
the energy of a partition $\sigma$ can be written as (denoting the set of edges of the graph by $V$)
\begin{eqnarray}
   H(\sigma) &=&  - \sum_{i,j \in V}\Big[\overbrace{a_{ij}A(i,j)-b_{ij}(1-A(i,j))}^{\text{internal edges}}\Big]\delta(\sigma_i,\sigma_j)
						 - \Big[\underbrace{c_{ij}A(i,j)-d_{ij}(1-A(i,j))}_{\text{external edges}}\Big](1-\delta(\sigma_i,\sigma_j))
\label{EnergyPartition}
\end{eqnarray}
where the minus sign preceding the summation symbol is a convention such that the optimal partition is characterised by a spin sate of minimal energy. \\
The terms of (\ref{EnergyPartition}) can be rearranged such that
\begin{eqnarray}
H(\sigma) &=& - H_0 - \sum_{i,j \in V}\Big[\alpha_{ij}A(i,j)-\beta_{ij}\Big]\delta(\sigma_i,\sigma_j)
\label{EnergyPartition2}
\end{eqnarray}
where $H_0 = \sum_{i,j \in V}\Big[-(c_{ij}+d_{ij})A(i,j)+d_{ij}\Big]$ is independent of the partition and can thus be removed from the fitness measure of the partition.\\

The values of the parameters 
\begin{eqnarray*}
    \alpha_{ij}&=& a_{ij}+b_{ij}+c_{ij}+d_{ij} \\
    \beta_{ij}&=&b_{ij}+d_{ij}
\end{eqnarray*}
depend on the null model the graph is compared to in order to compute the energy of the partition. The concept of null model is explained in what follows.

\subsection{Null model}

A crucial question to be raised when dealing with community detection is how to measure the quality of the partition obtained by some algorithm. Indeed, we do not know the \textit{a priori} community structure of most real networks and thus, we can not verify our results. This issue lead to the concept of \textit{null model}. \\

A null model is a graph that shares some features (to be chosen) with the original graph, but whose other features are essentially randomly determined. That randomness should avoid any community structure; by comparing the original graph with the null model, we can then prove the community structure of the first.\\

A major concern about null models is the selection of the features to be preserved. For instance, Reichardt and Borhnoldt chose an Erdös-Rényi graph as null model. In this null model, the number of nodes $N$ and the number of edges $m$ remain the same as in the original graph, but each edge is randomly (with constant probability) chosen to link two nodes. In other words, this null model is randomly taken in the set of all graphs which have $N$ nodes and $m$ edges. More formally, the probability that two nodes $i$ and $j$ are rewired, $p_{ij}$, is chosen in the Erdös-Rényi null model as
\begin{eqnarray}
p_{ij} = p
\end{eqnarray}
with $p$ constant.\\

Many other null models were designed either for generic or specific applications since the choice of a suitable null model can highly influence the performance measurements of an algorithm. In fact, two major drawbacks of the Erdös-Rényi model are known. On the one hand, it is hard to be generalised in the case of weighted graphs. On the other hand Erdös-Rényi graphs are known to exhibit a Poissonian degree distribution whereas real networks often have a degree distribution that follows a fat-tail power law.\\

\begin{figure}[h!]
\centering
    \includegraphics[scale=0.2]{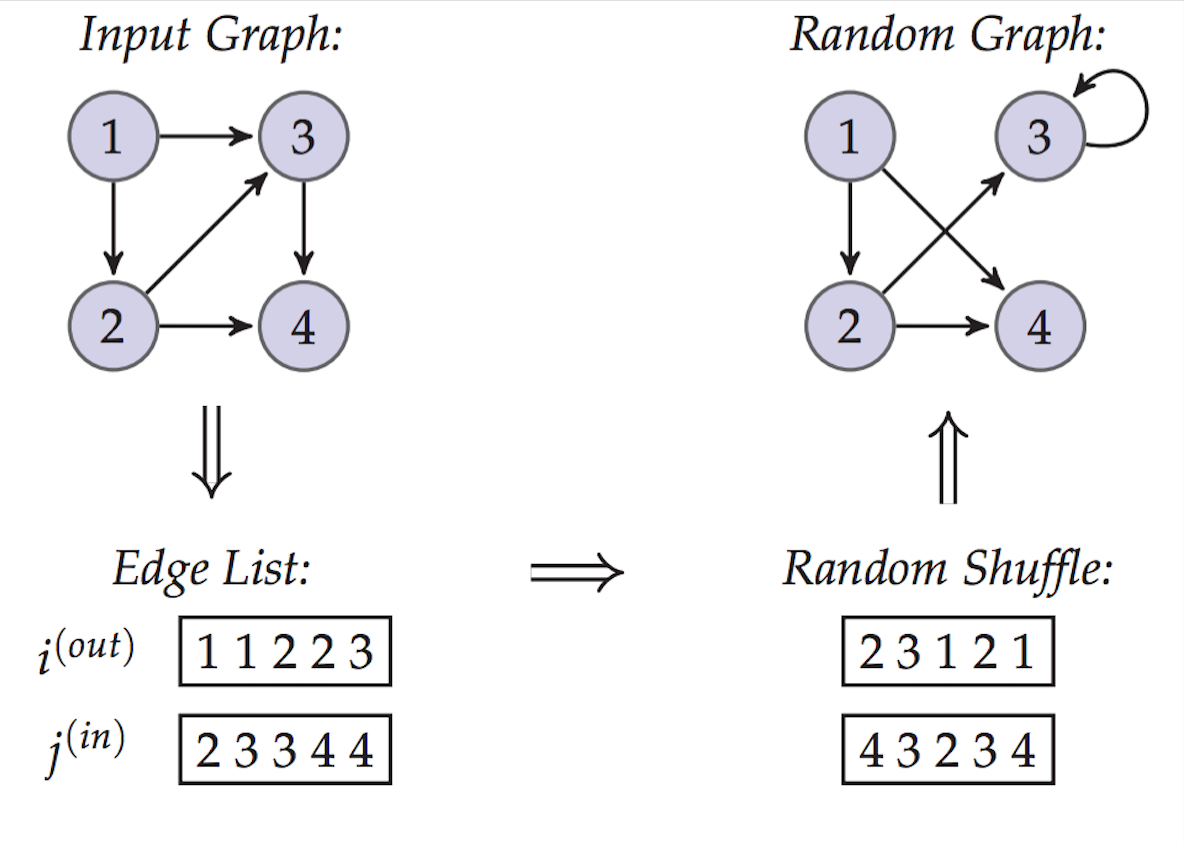}
    \caption[Configuration Null Model.]{Configuration Null Model. (Source: Browet, A. (2014) \textit{Algorithms for community and role detection in networks}. Ph.D. thesis in Engineering Sciences, Ecole Polytechnique de Louvain, Université Catholique de Louvain, Louvain-la-Neuve.)}
    \label{fig:configNullModel}
\end{figure}

Following this observation, Newman and Girvan introduced the so-called \textit{configuration null model} which is the basis of the modularity measure. The shared feature in the configuration null model is the degree distribution. To build this model from the original graph (in the directed case, for the sake of generality), the following steps must be followed: 
\begin{itemize}
    \item Build the edge list from the original graph. Let us define the vectors $i^{(out)}$ whose entries are the labels of the nodes having an outgoing edge (one entry for each edge), and $j^{(in)}$ whose entries are the labels of the corresponding nodes, i.e. the ones that have an incoming edge.
    \item Proceed to a random shuffle on both $i^{(out)}$ and $j^{(in)}$ vectors.
    \item The nodes of the random graph are rewired according to the shuffled edge list.
\end{itemize}
These steps are depicted in Figure (\ref{fig:configNullModel}). Intuitively, we can notice that this procedure does not change the degree distribution of the graph, as we wanted.\\
Based on this model, Newman and Girvan have proposed one of the most famous fitness measure for community detection: modularity.

\subsection{Modularity as fitness measure}

As said just above, the modularity is a fitness measure based on the configuration model producing a random graph with the same degree distribution  as the original network for which an optimal partition of the nodes into communities has to be found. \\

Given a partition $\sigma$, the modularity cost function compares the actual edge density within each community to the expected edge density in a random network using the configuration null model. It is given by
\begin{eqnarray}
Q(\sigma) &=& \sum_{i,j \in V}\Big[\frac{A(i,j)}{m} - \frac{k_i^{(out)}}{m}\frac{k_j^{(in)}}{m}\Big]\delta(\sigma_i,\sigma_j) \label{Modularity}
\end{eqnarray}
where $k_i^{(out)}$ (resp. $k_j^{(in)}$) is the number of outgoing (resp. incoming) edges of node $i$ (resp. $j$).\\
Up to scaling $\frac{1}{m}$, that does not modify the optimal partition, (\ref{Modularity}) can be expressed in the framework of the energy function, given by (\ref{EnergyPartition2}), with $\alpha_{ij}=1$ and $\beta_{ij}=p_{ij}$, i.e. the expected number of edge from $i$ to $j$.\\

Th modularity can easily be extended to the case of weighted networks:
\begin{equation*}
    Q_w(\sigma) = \frac{1}{m_w}\sum_{i,j \in V}\Big[W(i,j)-s_i^{(out)}\frac{s_j^{(in)}}{m_w}\Big]\delta(\sigma_i,\sigma_j)
\end{equation*}
where $W$ is the weighted adjacency matrix of the input graph and $s_i^{out}$ and $s_i^{in}$ are respectively the outgoing and incoming strengths  of nodes $i$. They are equal to the sum of the weights of the outgoing (respectively incoming) edges incidents to node $i$. (These notions are similar to those of outgoing and incoming degrees in unweighted graphs).
The total weight of the input graph, $m_w$, is
\begin{equation*}
    m_w = \sum_{i,j \in V} W(i,j) = \sum_{i\in V} s_i^{(out)} = \sum_{j\in V} s_j^{(in)}
\end{equation*}

The modularity score of a given partition belongs to the interval $[-1,1]$. It is equal to $0$ when the partition contains only one community and is in general negative for a partition with n communities, that is if every node defines a community. So, a partition with a high modularity score is supposed to be an accurate representation of the optimal partition of nodes into communities.\\
It can indeed be shown that  finding a community structure maximising the modularity function (and, more generally quality functions based on the energy of partitions) is a NP-hard problem. Hence, most of the community detection algorithms are greedy algorithms providing a good approximation of the optimal solution.\\ 

Modularity cost function is one of the most popular fitness measures and one of its biggest advantages is not to require to know beforehand the number of communities to extract. That is the quality function that is used in this report to determine a good partition of nodes into communities. But other fitness measures could be chosen too, in order to compensate some drawbacks related to the use of the modularity, which will not be discussed here. 


\section{The algorithm} \label{Algorithm}

Browet's community detection algorithm, also known as the Synchronised Louvain Method, is summarised below by its description using Algorithm \ref{AlgoBrowet}. 
\begin{algorithm}
\caption{Synchronised Louvain Method}
\label{AlgoBrowet}
\begin{algorithmic}
\STATE \textbf{Input}: a graph $G(V,E)$ 
\STATE \textbf{Output}: a community partition matrix $C \in \mathbb{R}^{k \times n}$
\STATE Initialize $C=I_n, \;\; C_t=0, \;\; G_t=G$
\STATE $\;\;\;\;\;\;\;\;\;\;\;\;\;\;\;\;\;\;\;\;\;\;\rhd$ $C$ final partition, $C_t$ partition at level $t$ for $G_t$ 
\WHILE {$C_t\neq I$}
\STATE $C_t \leftarrow \text{ASSIGN} (G_t)$
\STATE $C_t \leftarrow \text{POSITIVE} (C_t,G_t)$
\WHILE{$\exists \; i \in V_t, \; c \in C_t\; \text{with}\; \Delta H(c_i \rightarrow i \rightarrow c) > 0$}
\STATE $C_t \leftarrow \text{MAXIMAL}(C_t,G_t)$
\STATE $C_T \leftarrow \text{POSITIVE}(C_t,G_t)$
\ENDWHILE
\STATE $G_t \leftarrow \text{AGGREGATE}(G_t)$
\STATE $C=C_tC$
\ENDWHILE
\end{algorithmic}
\end{algorithm}
\\
In this report, we focus on the most important parts of the algorithm for the project, i.e. for the parallel implementation of Browet's algorithm. But it is described in more details in Arnaud Browet's thesis \cite{ThesisBrowet}. 

We will explain in details the important parts of the algorithm in the following sections. The pseudocodes of these main steps are shown in the appendix (see \ref{pseudocodes}). 

\subsection{ASSIGN (best neighbour)}
\begin{figure}[h!]
    \centering
    \includegraphics[scale=0.6]{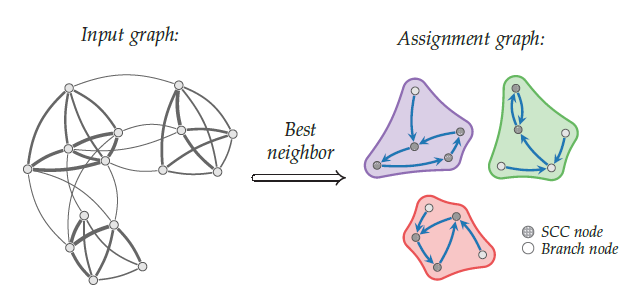}
    \caption[Assignation and extraction of connected components steps.]{Assignation and extraction of connected components steps. (Source: Browet, A. (2014) \textit{Algorithms for community and role detection in networks}. Ph.D. thesis in Engineering Sciences, Ecole Polytechnique de Louvain, Université Catholique de Louvain, Louvain-la-Neuve.)}
    \label{ASSIGN}
\end{figure}

The first step of the algorithm consists in initialising the community structure by assigning each node of the graph to its best neighbour. \\
\\
In order to do so, the algorithm computes for each of the neighbours of the current node the gain in modularity obtained by assigning the current node i to the community formed by this neighbour j (i.e. $\Delta H(i \rightarrow \{j\})$). The best neighbour of the node is the one for which the modularity gain is the highest. (If all the neighbours of a node are such that this gain is negative, the node is self-assigned, in which case the gain is equal to $0$).\\
\\
At the end of the assignment step, we obtain a set of directed subgraphs spanning the input graph, which is called the assignment graph, with one directed edge going out from each node to its best neighbour, as illustrated in Figure \ref{ASSIGN}. 

The complexity of the assignment step is $\mathcal{O}(m)$, where $m$ is the number of edges in the input graph. 

\subsection{COMPONENTS (connected components)}
The next step is to extract the connected components of the assignment graph, which are defined as the communities. For example, the assignment graph in Figure \ref{ASSIGN} consists of 3 communities. As illustrated on the same figure, each community contains one and only one directed cycle, the strongly connected component (SCC). \\

Given that the assignation of each node to its best neighbour is independent from the assignation of the other nodes, the assignation step might result in communities in which the presence of some nodes decreases the value of the cost function. \\
In the following steps, two types of corrections will thus be applied to the assignment graph in order to improve the community partition, by updating some node assignments.  

\subsection{POSITIVE (split)}
The positive correction ensures that each node has a positive contribution to its community in terms of the cost function. \\
The local gain of a node is defined as the modularity gain to assign the node to its current community. For each community containing at least one node with a negative local gain, the algorithm searches for the optimal (i.e. providing the greatest gain of modularity) bisection (split) of these communities. \\

\begin{figure}[h!]
   \begin{minipage}[c]{.46\linewidth}
      \includegraphics[scale=0.4]{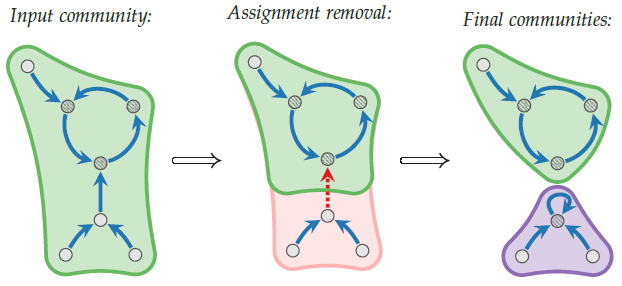}
      \caption[Positive correction in a branch.]{Positive correction in a branch. (Source: Browet, A. (2014) \textit{Algorithms for community and role detection in networks}. Ph.D. thesis in Engineering Sciences, Ecole Polytechnique de Louvain, Université Catholique de Louvain, Louvain-la-Neuve.)}
      \label{PositiveCorrectionBranch}
    \end{minipage} \hfill
   \begin{minipage}[c]{.46\linewidth}
      \includegraphics[scale=0.4]{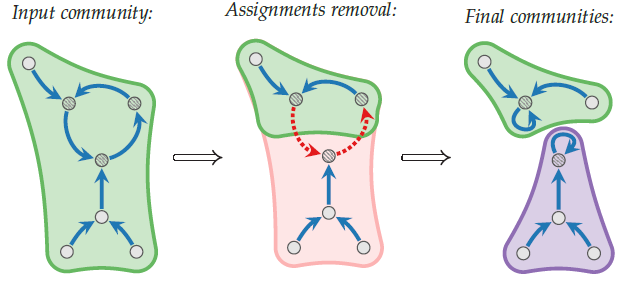}
      \caption[Positive correction in the SCC.]{Positive correction in the SCC. (Source: Browet, A. (2014) \textit{Algorithms for community and role detection in networks}. Ph.D. thesis in Engineering Sciences, Ecole Polytechnique de Louvain, Université Catholique de Louvain, Louvain-la-Neuve.)}
      \label{PositiveCorrectionSCC}
   \end{minipage}
\end{figure}

A positive correction is made by updating at most two node assignments. Indeed, two kinds of bisections can be considered. Either we remove the assignment of a node in a branch of the community (as shown on Figure \ref{PositiveCorrectionBranch}) or two assignments within the SCC (as illustrated in Figure \ref{PositiveCorrectionSCC}). The nodes whose assignments are removed are self-assigned. 

Positive corrections are applied till there is no node with a negative local gain anymore. \\

The complexity of the Positive Correction step is dominated by a linear complexity in the number of nodes within the branches of a community. 

\subsection{MAXIMAL (merge)}
\label{Maximal}
Once all possible positive corrections have been made, maximal corrections are applied to the assignment graph by allowing some nodes to switch from one community to another in order to further optimise the cost function and thus the community partition. \\

\begin{figure}[h!]
\centerline{
   \begin{minipage}[c]{.46\linewidth}
      \includegraphics[scale=0.47]{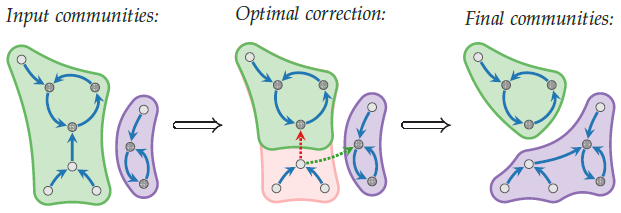}
      \caption[Maximal correction in a branch.]{Maximal correction in a branch. (Source: Browet, A. (2014) \textit{Algorithms for community and role detection in networks}. Ph.D. thesis in Engineering Sciences, Ecole Polytechnique de Louvain, Université Catholique de Louvain, Louvain-la-Neuve.)}
      \label{MaximalCorrectionBranch}
   \end{minipage} \hfill
   \begin{minipage}[c]{.46\linewidth}
      \includegraphics[scale=0.47]{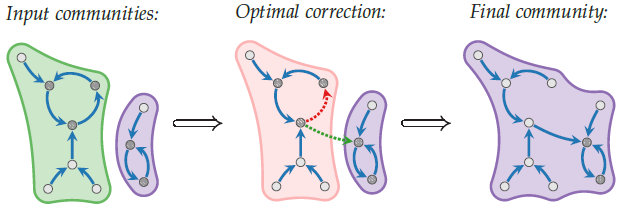}
      \caption[Maximal correction in the SCC.]{Maximal correction in the SCC. (Source: Browet, A. (2014) \textit{Algorithms for community and role detection in networks}. Ph.D. thesis in Engineering Sciences, Ecole Polytechnique de Louvain, Université Catholique de Louvain, Louvain-la-Neuve.)}
      \label{MaximalCorrectionSCC}
   \end{minipage}}
\end{figure}

While positive corrections induce an increase in the number of communities by splitting them, the maximal corrections tend to decrease this number by merging some communities (we will come back on this later on). \\
One should notice that, when a node is switching from a community to another one, all the nodes in its tail (the set of nodes such that there exists a directed path leading from them to the node in the assignment graph) are also switching to its new community, as illustrated in Figure \ref{MaximalCorrectionBranch}. \\
In order to decide to which community a node will be assigned, the algorithm computes the gains (given the current partition) obtained by assigning the node and the nodes of its tail to each of the communities in which the node has a neighbour in the input graph (thus not in the assignment graph). Then, the node and its tail are assigned to the community providing the largest non negative gain (it can be its current community). To ensure the convergence of the algorithm, each maximal correction is accepted with a certain probability (see \cite{ThesisBrowet} for more details). \\

A maximal correction results in a merge of two communities if the node that is switched to another community is part of the SCC (as shown on Figure \ref{MaximalCorrectionSCC}) because its tail is the entire community. But this is not the case if the node belongs to a branch of the community (see Figure \ref{MaximalCorrectionBranch}).

The complexity of the Maximal Corrections step is $\mathcal{O}(m)$. \\

Positive and maximal corrections are alternatively applied till convergence is reached.\\

\subsection{AGGREGATE (aggregation of communities )}
\begin{figure}[h!]
    \centerline{
    \includegraphics[scale=0.65]{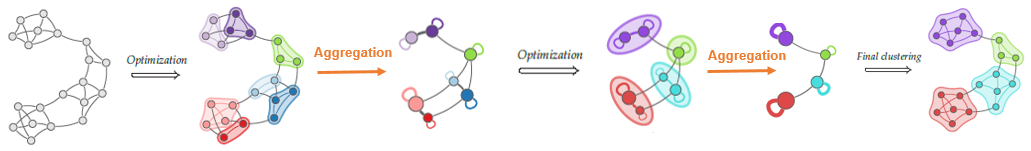}}
    \caption[Aggregation step.]{Aggregation step. (Source: Browet, A. (2014) \textit{Algorithms for community and role detection in networks}. Ph.D. thesis in Engineering Sciences, Ecole Polytechnique de Louvain, Université Catholique de Louvain, Louvain-la-Neuve.)}
    \label{Aggregation}
\end{figure}

When no correction providing a strictly positive gain can be done, the community graph is collapsed. All the nodes belonging to the same community are aggregated into one "super node". Then the same procedure (Assign, Components, Positive, Maximal, Positive, ..., Aggregate) is applied to the aggregated graph, which provides another hierarchical level of clustering. This is illustrated in Figure \ref{Aggregation}. 
The algorithm stops when it can not extract any community structure in the last aggregated graph, i.e. when the assignment of any "super node" to any other "super node" results in a negative gain. \\
\\

The total complexity of the algorithm can not be computed because it depends on the number of positive and maximal corrections, which can not be evaluated. 


\section{Parallel implementation}
\subsection{What is parallel computing?}
\begin{figure}[h!]
   \begin{minipage}[c]{.46\linewidth}
    \includegraphics[scale=0.5]{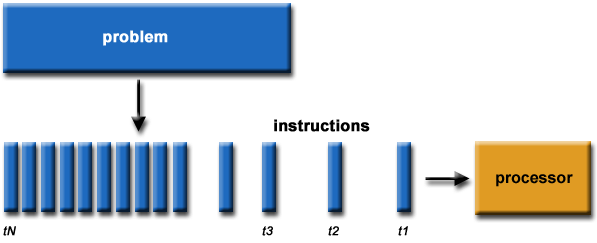}
    \caption[Sequential computing.]{Sequential computing. (Source: Barney, B. (2016) \textit{Introduction to Parallel Computing}. \url{https://computing.llnl.gov/tutorials/parallel\_comp/}. )}
    \label{SequentialComputation}
   \end{minipage} \hfill
   \begin{minipage}[c]{.46\linewidth}
    \includegraphics[scale=0.4]{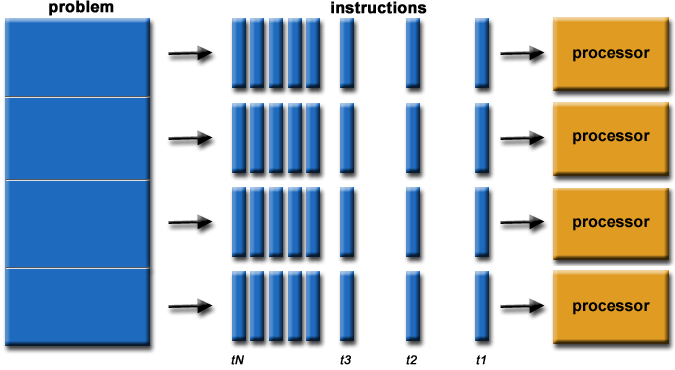}
    \caption[Parallel computing.]{Parallel computing. (Source: Barney, B. (2016) \textit{Introduction to Parallel Computing}. \url{https://computing.llnl.gov/tutorials/parallel\_comp/}.)}
    \label{ParallelComputation}
   \end{minipage}
\end{figure}

As explained in \cite{ParallelComputing}, most of the programs we use today are written to be sequentially executed. This means that the program is broken into a discrete series of instructions who are executed one after another, on a single processor, as illustrated in Figure \ref{SequentialComputation}. 

Parallel computing takes advantage of the increasing number of processor cores in computers. The main idea of parallel computing is to distribute the work to be done by a program between several cores. It follows in a significant computing time saving.\\
The problem is broken into discrete pieces of work that can be solved independently, each part being handled by one core. A part consists in a sequence of instructions that are solved one after another by a same core. An overall control mechanism is employed to divide the work and to ensure the coordination and synchronisation between the cores, as shown in Figure \ref{ParallelComputation}.\\ 
The results of the program when executed in parallel must be the same that when sequentially executed. \\

Not every algorithm can be parallelised: indeed, it has to be written such that the instructions can be executed independently. Let us consider algorithms \ref{ExParallel}  and \ref{ExNonParallel}. Both are computing the same thing: the index of the elements in the vector \textit{count}. But they cannot be both parallelised. Indeed, the for-loop in algorithm \ref{ExParallel} can be parallelised, i.e. that, if we have $n$ cores, $n$ iterations of the loop can be executed simultaneously, one by each core, because each iteration is independent from the other ones. But this is not the case of the loop in algorithm \ref{ExNonParallel} because each iteration depends on the previous one: you cannot compute $\text{count}(3)$ if the value of $\text{count}(2)$ has not been computed yet given that there is a temporal dependence between the iterations. 

\begin{minipage}[t]{7.5cm}
  \vspace{0pt}  
  \begin{algorithm}[H]
    \caption{Parallelisable Program}
    \label{ExParallel}
    \begin{algorithmic}
     \STATE $\text{count}=\text{zeros}(1,100)$
    \FOR{$i=1...100$}
    \STATE $ \text{count}(i)=i$
    \ENDFOR
    \STATE
    \end{algorithmic}
  \end{algorithm}
\end{minipage}%
\begin{minipage}[t]{7.5cm}
  \vspace{0pt}
  \begin{algorithm}[H]
  \begin{algorithmic}
    \caption{Non Parallelisable Program}
    \label{ExNonParallel}
    \STATE $\text{count}=\text{zeros}(1,100)$
    \STATE $\text{count}(1)=1$
    \FOR{$i=2...100$}
    \STATE $ \text{count}(i)=\text{count}(i-1)+1$
    \ENDFOR
    \end{algorithmic}
  \end{algorithm}
\end{minipage}

\subsection{Parallel implementation of Browet's algorithm}
Let us now see what parts of the Synchronised Louvain Method are parallelisable, without modifying Browet's code. Among the four main steps detailed in section \ref{Algorithm}, three of them can be parallelised: the assignation step and the positive and maximal corrections.\\

\begin{figure}[h!]
   \centering
   \includegraphics[scale=0.6]{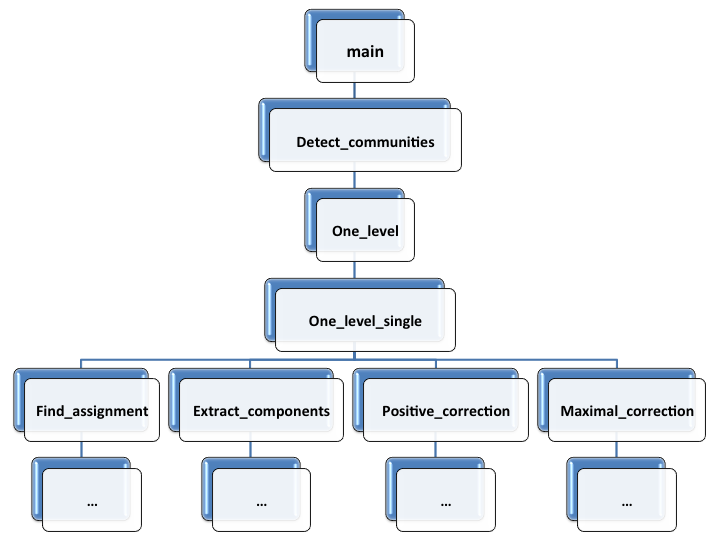}
   \caption{Hierarchical organisation of the code to be parallelised (call-graph)}
   \label{fig:callgraph}
\end{figure}
We can see on Figure \ref{fig:callgraph}, obtained by code profiling, that the highest level we can parallelise contains four functions: \texttt{find\_assignment()}, \texttt{extract\_components()}, \texttt{positive\_correction()} and \texttt{maximal\_correction()} (the first four levels are not). We implemented a parallel version for all of them, except for the \textbf{connected components extraction} step. Indeed, this step was not directly parallelisable. However, the execution time spent in this function is negligible compared to the other functions. In fact, this step would have required a certain amount of computation time, but in this case not. Given that the complexity of this step is growing with the number of edges in the graph and that, after the assignation step, this number of edges is drastically reduced ($m=n$), the computation time is really small. Therefore, we did not to implement a parallel version of \texttt{extract\_components()}.\\

The \textbf{assignation step} of the algorithm can be run in parallel given that the initial assignment of each node is independent of the one of the other nodes. The way the best neighbour of a node is chosen is such that the algorithm only needs the input graph to compute it: the best neighbour of a node does not influence the "computation" of the best neighbour of another node. This allows us to use a parallel processor architecture: each core can handle a set of nodes independently. 

Assigning each node to another node allows the \textbf{corrections} to be made in parallel (synchronously). Indeed, if a node has to be switched from a community to another one, we simply have to change its assignation to a node of its new community. This can be done independently of the assignation of the other nodes in the network (this reassignation does not modify the assignation of the other nodes). The correction on any node is thus independent from the corrections on other nodes. \\
\\
The \textbf{positive correction} of a community does not depend on the state of the other communities so this step is parallelisable: each split can be assigned to a different core. Indeed, the computation of the local gain of a node only depends on the structure of the community  it belongs to and the search for an optimal bisection of a community is only dependent on the structure of this community.\\
\\
To apply a \textbf{maximal correction}, the algorithm does not use any other information than the current community distribution in order to compute the switching gains and the community structure is updated after all the new assignments have been computed. Here again, communities can be "corrected" simultaneously, each core handling a community. 

\subsection{OpenMP}
\begin{figure}[h!]
\centerline{
   \begin{minipage}[c]{.4\linewidth}
    \includegraphics[scale=0.5]{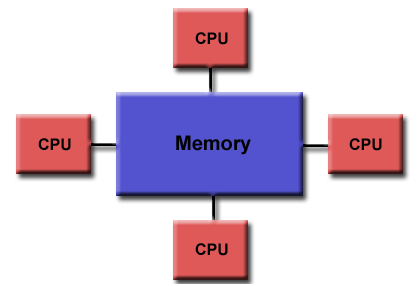}
    \caption[Shared memory.]{Shared memory. (Source: Barney, B. (2015) \textit{OpenMP}.  \url{https://computing.llnl.gov/tutorials/openMP/})}
    \label{SharedMemory}
   \end{minipage} \hfill
   \begin{minipage}[c]{.55\linewidth}
    \includegraphics[scale=0.5]{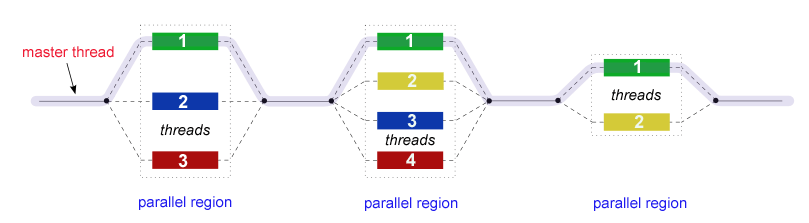}
    \caption[Fork-join model.]{Fork-join model. (Source: Barney, B. (2015) \textit{OpenMP}. \url{https://computing.llnl.gov/tutorials/openMP/})}
    \label{Fork-Join}
   \end{minipage}}
\end{figure}

In order to parallelise the Synchronised Louvain Method (which was coded in C++), we used OpenMP \cite{OpenMP}, which is an Application Program Interface (API) that may be used to direct multi-threaded, shared memory parallelism. In the case of shared memory parallelism, all cores share the same memory (see Figure \ref{SharedMemory}), contrary to distributed memory parallel systems where each core has its own memory. 

OpenMP uses the fork-join model (shown in Figure \ref{Fork-Join}) of parallel computing. The program begins as a single process: the master thread. The instructions are executed sequentially by the master thread till it encounters a parallel region. At this point, it creates parallel threads (fork), which exist only in the parallel region. Then, the instructions in the parallel region are simultaneously executed by the parallel threads. When all the statements in the parallel region have been executed, they synchronise, terminate and only the master thread does not cease to exist (join) till the next parallel region. All the threads must have finished before the following sequential region is executed by the master thread.\\

Given that all the threads share the same memory, they all can read and write the data simultaneously and all changes made are visible for all threads. This may raise problems in some cases when different threads try to modify the same variable simultaneously. This is why an important task in the parallelisation of Browet's algorithm has been to determine which variables (declared before the parallel region but used inside it) were public (all threads have access to them) and which are private (each thread has a copy of this variable and is the only one to have access to it).\\
If we take a closer look at algorithm \ref{ExParallelPrivatePublic} (which computes the index of the elements in the vector \textit{count)}, one can see that the variable \textit{iterations} has to be public because it counts the total number of iterations of the for-loop. But the variable \textit{position} has to be private. This can be illustrated by this simple example. If a thread is handling the iteration (i=2), \textit{position} and thus \textit{count(2)} are equal to 2. If another thread is handling iteration (i=4), \textit{position} and thus \textit{count(4)} are equal to 4. If both threads are executing these operations at the same time, it could be possible that \textit{count(2)}=4 or that \textit{count(4)}=2, which is not correct given that the results of the sequential and of the parallel executions of the program are not the same anymore. \\
  \begin{algorithm}[H]
    \caption{}
    \label{ExParallelPrivatePublic}
    \begin{algorithmic}
     \STATE $\text{count}=\text{zeros}(1,100)$
     \STATE $\text{iterations} = 0$
     \STATE $\text{position}$
    \FOR{$i=1...100$}
    \STATE $\text{position}=i$
    \STATE $ \text{count}(i)=\text{position}$
    \STATE $\text{iterations}++$
    \ENDFOR
    \end{algorithmic}
  \end{algorithm}

A region is parallelised by inserting directives in the code, for instance
\begin{center}
\#pragma omp parallel for private(position) schedule(dynamic)\\
\{... for loop ...\}
\end{center}
In this example, the "for" means that the parallel region is a for loop and the "private(position)" that the variable \textit{position} is private. The "schedule" clause is used to define how the iterations of the loop are divided among the threads.
If one specifies that they have to be divided on a static way, each thread will execute approximately $\frac{n}{t}$ (where $n$ stands for the number of iterations and $t$ for the number of threads) iterations of the for loop. This is not optimal when all the iterations do not take more or less the same time to be executed. If this is the case, it is a better idea to use the option "dynamic" that ensures that the work is distributed more evenly between the threads (when a thread has finished to execute an iteration, another ond is dynamically assigned to it).   \\

Another important task in the parallel implementation of Browet's algorithm has been to determine if, within a parallel region, a certain region was critical. A  critical region is a sequence of instructions that has to be executed by one thread at once. The corresponding directive is
\begin{center}
\#pragma omp critical\\
\{... critical region ...\}
\end{center}

To parallelise the Synchronised Louvain Method, we added omp directives in the code of Arnaud Browet (in the functions ASSIGN, POSITIVE and MAXIMAL). In order to know where to add them, we had to understand the above mentioned functions, try to see if there were temporal dependencies, if the variables had to be private or public, if some parts of the parallel regions had to be critical, and so on.


\section{Results}
\subsection{Benchmark graphs}
\begin{figure}[H]
    \centerline{
    \includegraphics[scale = 0.55]{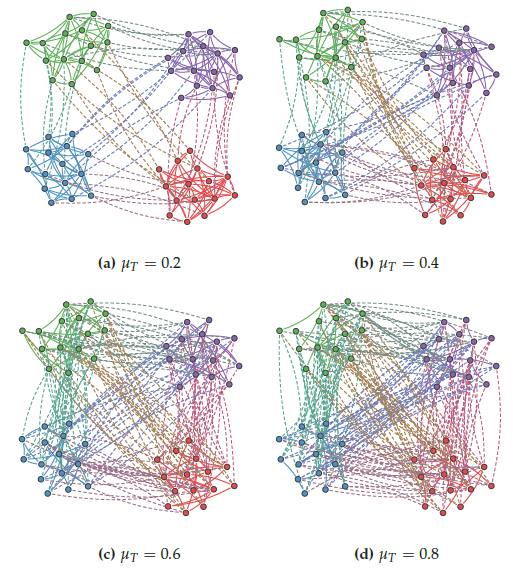}}
    \caption[Influence of mixing parameters.]{Influence of mixing parameters. (Source: Browet, A. (2014) \textit{Algorithms for community and role detection in networks}. Ph.D. thesis in Engineering Sciences, Ecole Polytechnique de Louvain, Université Catholique de Louvain, Louvain-la-Neuve.)}
    \label{fig:mix_param}
\end{figure}

To test the performances of the parallelised algorithm, we generated benchmark graphs using the so-called LFR benchmark algorithm \cite{LFR}. This kind of benchmark graphs have the advantage to get closer to "real-world" networks' topology than graphs generated by other methods. All the graphs we used were weighted and directed because the Synchronised Louvain Method has been written for such graphs. The needed parameters and the ranges we used for each one of them are listed in Table \ref{tab::LFRparam}.\\
\begin{table}[H]
\centerline{
\begin{tabular}{|c|c|c|}
\hline \textbf{Parameter} & \textbf{Meaning} & \textbf{Tested values} \\ 
\hline
\hline $N$ & Number of nodes & $ie3, ie4, ie5, i \in \{1,2,4,6,8\}, 1e6$ \\ 
\hline $k$ & Mean degree of nodes & $50$ \\ 
\hline $kmax$ & Maximum degree of nodes & $100$ \\ 
\hline $(\mu_{\tau},\mu_w)$ & Mixing parameters (topology and weights) & $(0.2,0.1), (0.5,0.4), (0.8,0.7)$ \\ 
\hline 
\end{tabular}}
\caption{Used parameters for the generation of benchmark graphs.}
\label{tab::LFRparam}
\end{table}
The mean and maximum degrees of nodes were chosen constant in order to get the same edges density for each benchmark graph. The mixing parameters ($\in [0,1]$) influence the way the communities are mixed between each other. The lower they are, the more the communities are distinct and easy to detect (Figure \ref{fig:mix_param}). For example, the parameter $\mu_{\tau} = 0.2$ (resp. $\mu_w = 0.1$) means that the edges density (resp. the weights of the edges) \textit{inside} each community is (resp. are) way higher than the one(s) \textit{between} communities. To avoid "ill-posed" graphs, we always chose $\mu_{\tau} > \mu_w$.\\
For each set of parameters, we generated two different graphs and computed the mean results. The time results in the following parts are given using "wall-clock" time, i.e. the time that the user really waits for the execution of the algorithm, from the beginning to the end.

\subsection{Computation time}
\begin{figure}[H]
  \centerline{
  \subfloat[]{\label{TNa}\includegraphics[scale=0.6]{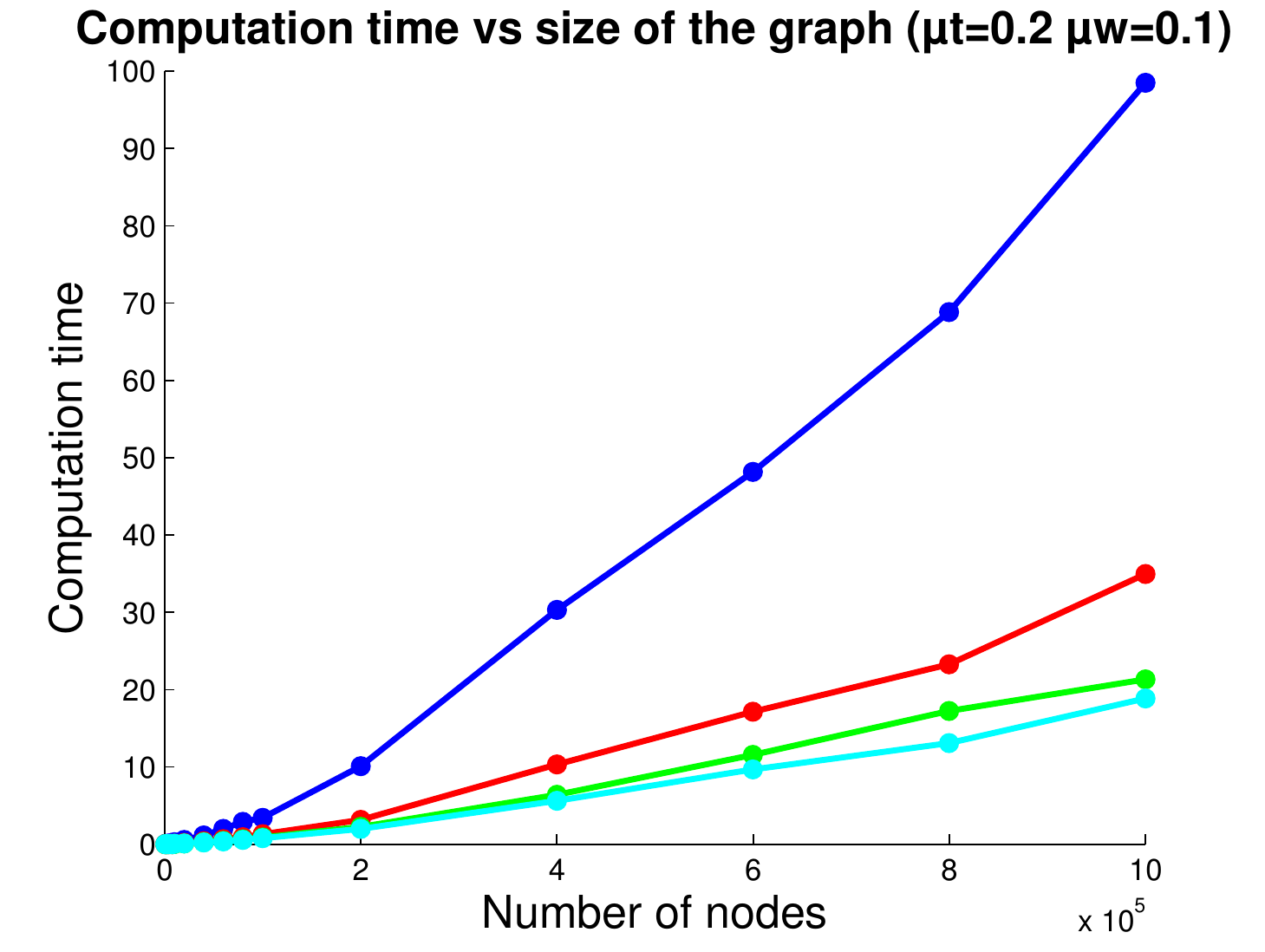}}
  \hspace{5pt}
  \subfloat[]{\label{TNb}\includegraphics[scale=0.6]{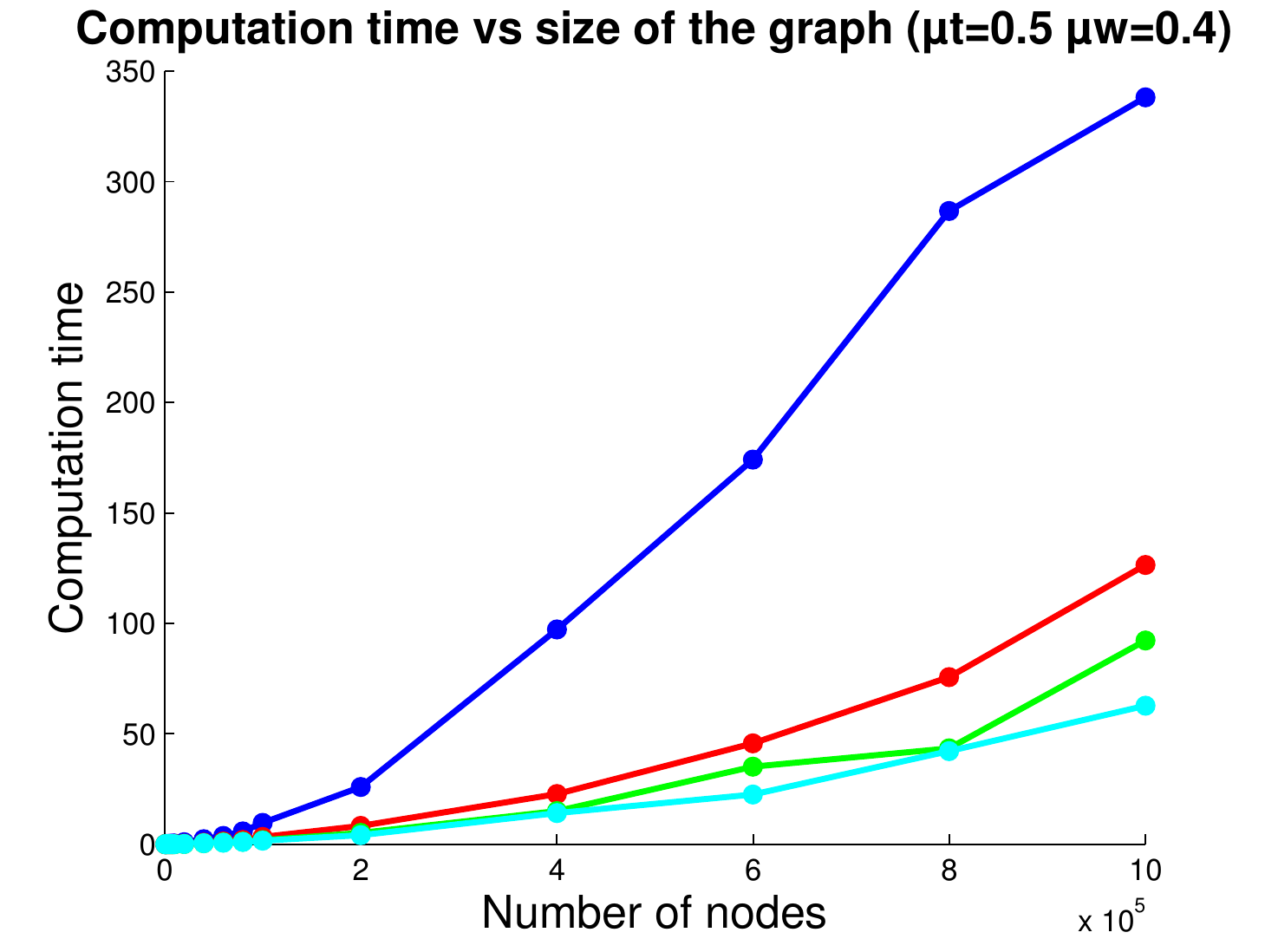}}}
  \hspace{5pt}
  \centerline{
  \subfloat[]{\label{TNc}\includegraphics[scale=0.6]{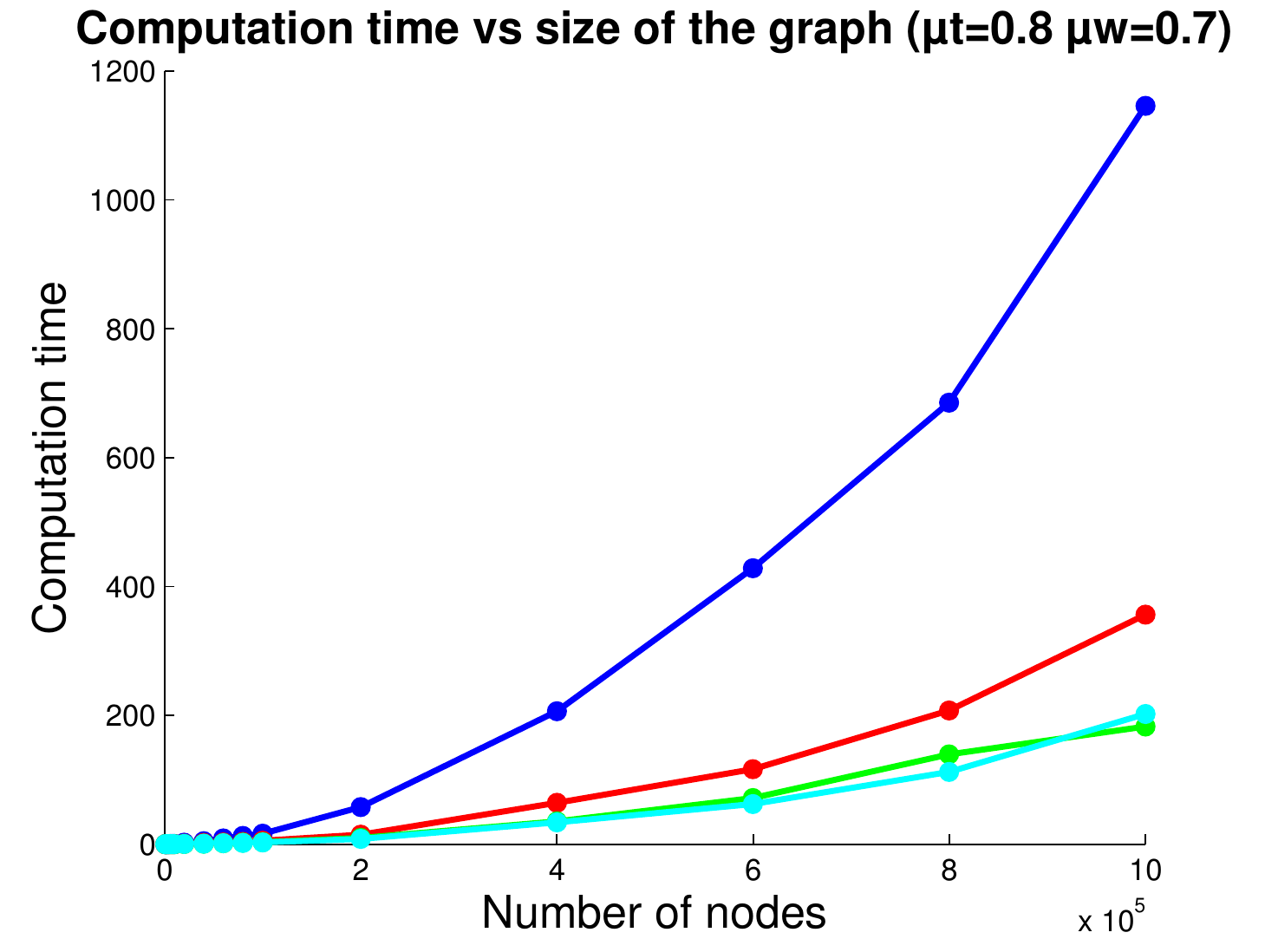}}
    \hspace{5pt}
    \hspace{2.5cm}
  \subfloat{\label{TNd}\includegraphics[scale=1]{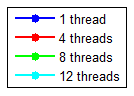}}}
  \caption{Computation time versus number of nodes.}
  \label{TN}
\end{figure}

Computation time with respect to the number of nodes and depending on the number of threads is represented in Figure \ref{TN}. Three trends seem to appear on the Figures \ref{TNa}-\ref{TNc}. \\

The first one is that the computation time clearly decreases with the number of threads, as expected. \\
Secondly, by looking at these figures we can see that it seems that using the parallel implementation of Browet's algorithm is really interesting for large enough graphs. Indeed, for graphs with less than $2.10^3$ nodes, there is no real difference in terms of computation time between sequential and parallel executions. But, the higher the size of the graph, the greater the difference between sequential and parallel (12 threads) times, and thus the higher the gain in computation time. \\
Finally, these figures illustrate well the influence of the mixing parameters on the computation time. The higher they are, the higher the execution time. Indeed, high values for the mixing parameters mean that the communities are less distinct and thus that more corrections are needed before convergence is reached. And corrections are the most expensive functions in terms of computation time. On the contrary, graphs with more distinct communities need less corrections before reaching convergence and thus the computation time is smaller. 

\subsection{Speedup}
\subsubsection{Expected and real speedup}
\label{Amdahl}
\begin{figure}[H]
\centerline{
    \includegraphics[scale=0.45]{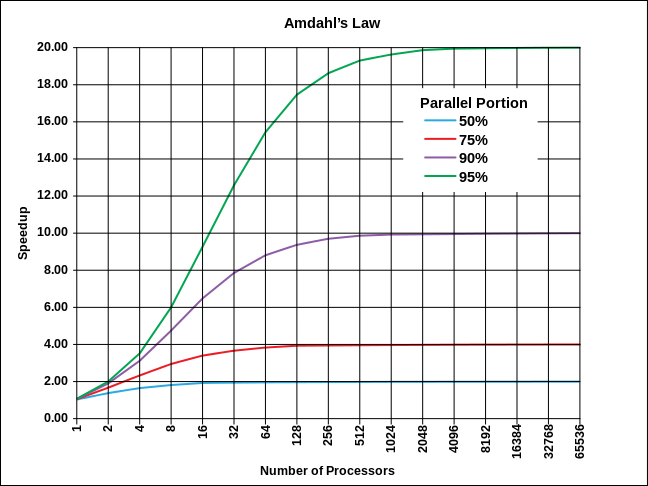}}
    \caption[Amdahl's law.]{Amdahl's law. (Source: Wikipedia contributors, \textit{Amdahl's law, Wikipedia, The Free Encyclopedia}, \url{https://en.wikipedia.org/w/index.php?title=Amdahl\%27s\_law&oldid=716179061} (accessed May 5, 2016).)}
    \label{fig:amlaw}
\end{figure}

To evaluate the theoretical maximum speedup we can obtain from our parallel implementation, one way is to refer to Amdahl's law (see \cite{Amdahl}):
\begin{eqnarray}
S &=& \frac{1}{(1-P)+\frac{P}{N}} \label{eq::amlaw}
\end{eqnarray}
where 
\begin{itemize}
\item $S$ is the maximum expected speedup
\item $P$ is the portion of parallelised code in terms of wall-clock \textit{sequential} execution time: $$P = \frac{\text{Fraction of time spent  in parallelised parts}}{\text{Total execution time}}$$
\item $N$ is the number of threads being used
\end{itemize}
The theoretical speedup predicted by Amdahl's law is illustrated in Figure \ref{fig:amlaw}. However, we can notice that the speedup of any parallel program is always bounded given that the execution time of the non parallelisable part of the program can not be reduced.\\

\begin{figure}[H]
\centerline{
    \includegraphics[scale=0.7]{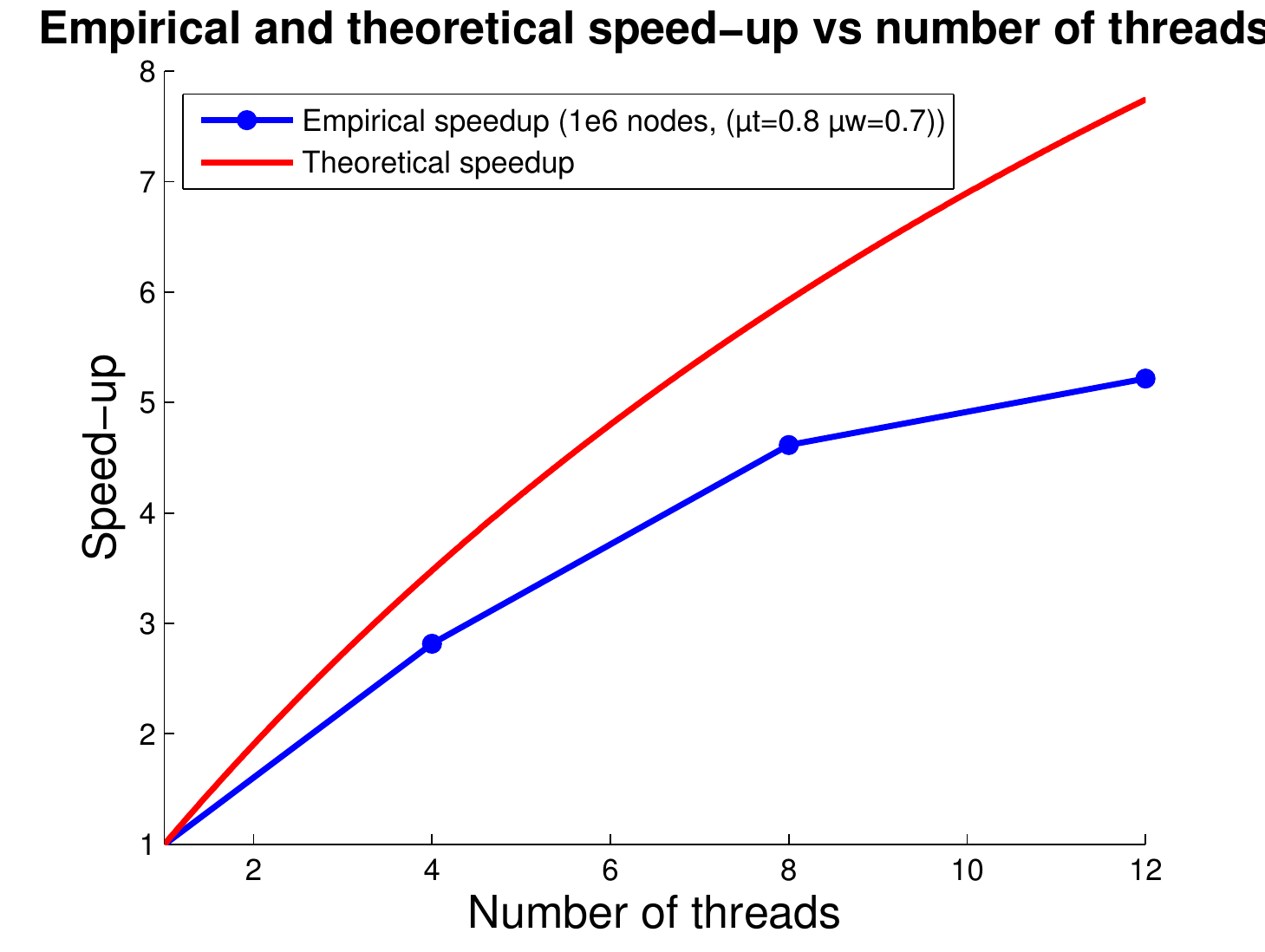}}
    \caption{Theoretical (for $P=95\%$) and empirical speedup.}
    \label{SpeedUps_Th_Emp}
\end{figure}

Even if Amdahl's law is often used to predict the potential speedup, we need to be careful with this result. Indeed, this law is purely theoretical and does not take into account some factors. One of them, which is critical, is the overhead phenomenon.\\
The overhead sums up the amount of external processing time needed for the execution of a code. In the case of a parallel program, the overhead is increased by the installation of parallelism, the synchronisation and communication between threads, and so on. In fact, we shall see that, due to increased overhead, our parallel implementation is only useful for large graphs, when the added overhead is sufficiently compensated by the acceleration gained by parallelism. Therefore, we can deduce that the expected speedup is actually lower than predicted by Amdahl's law, as we can see on Figure \ref{SpeedUps_Th_Emp} where we arbitrarily used $P = 0.95$. To compute the exact $P$ for our implementation, we could have used a code profiler, but this was not possible due to technical reasons.. 

\subsubsection{Influence of size, mixing parameters and number of threads}
\begin{figure}[H]
  \centerline{
  \subfloat[]{\label{SNa}\includegraphics[scale=0.6]{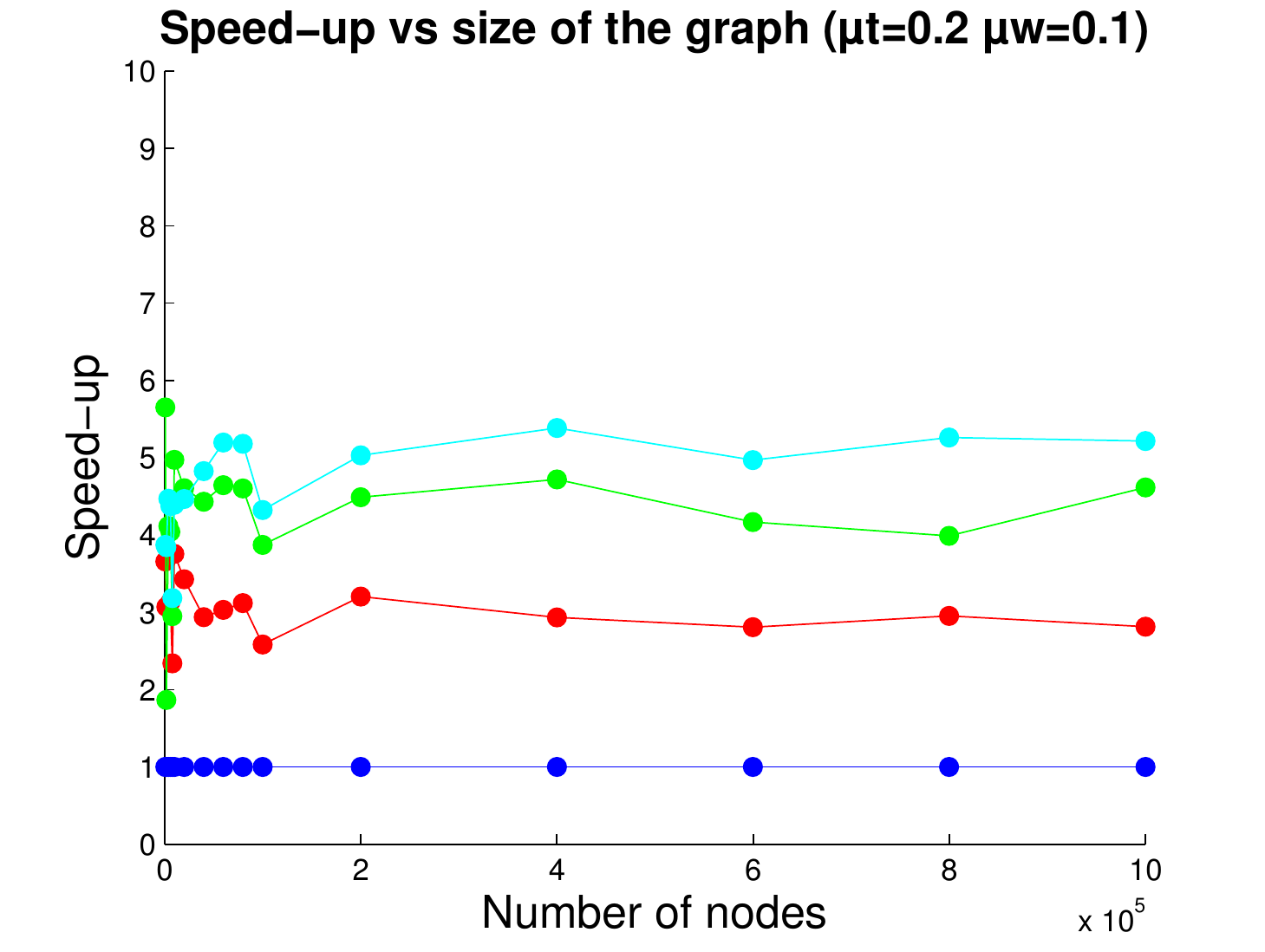}}
  \hspace{5pt}
  \subfloat[]{\label{SNb}\includegraphics[scale=0.6]{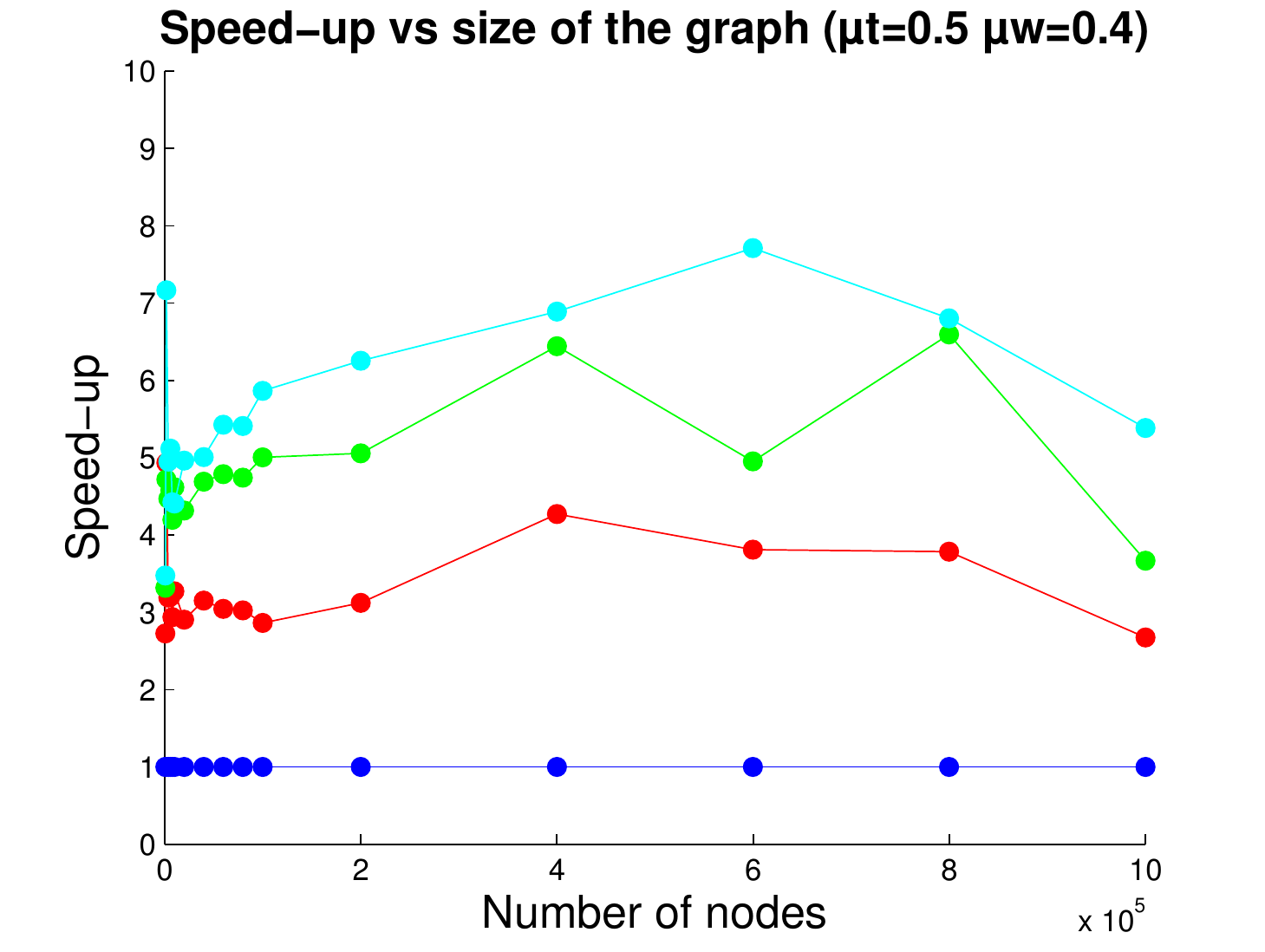}}}
  \hspace{5pt}
  \centerline{
  \subfloat[]{\label{SNc}\includegraphics[scale=0.6]{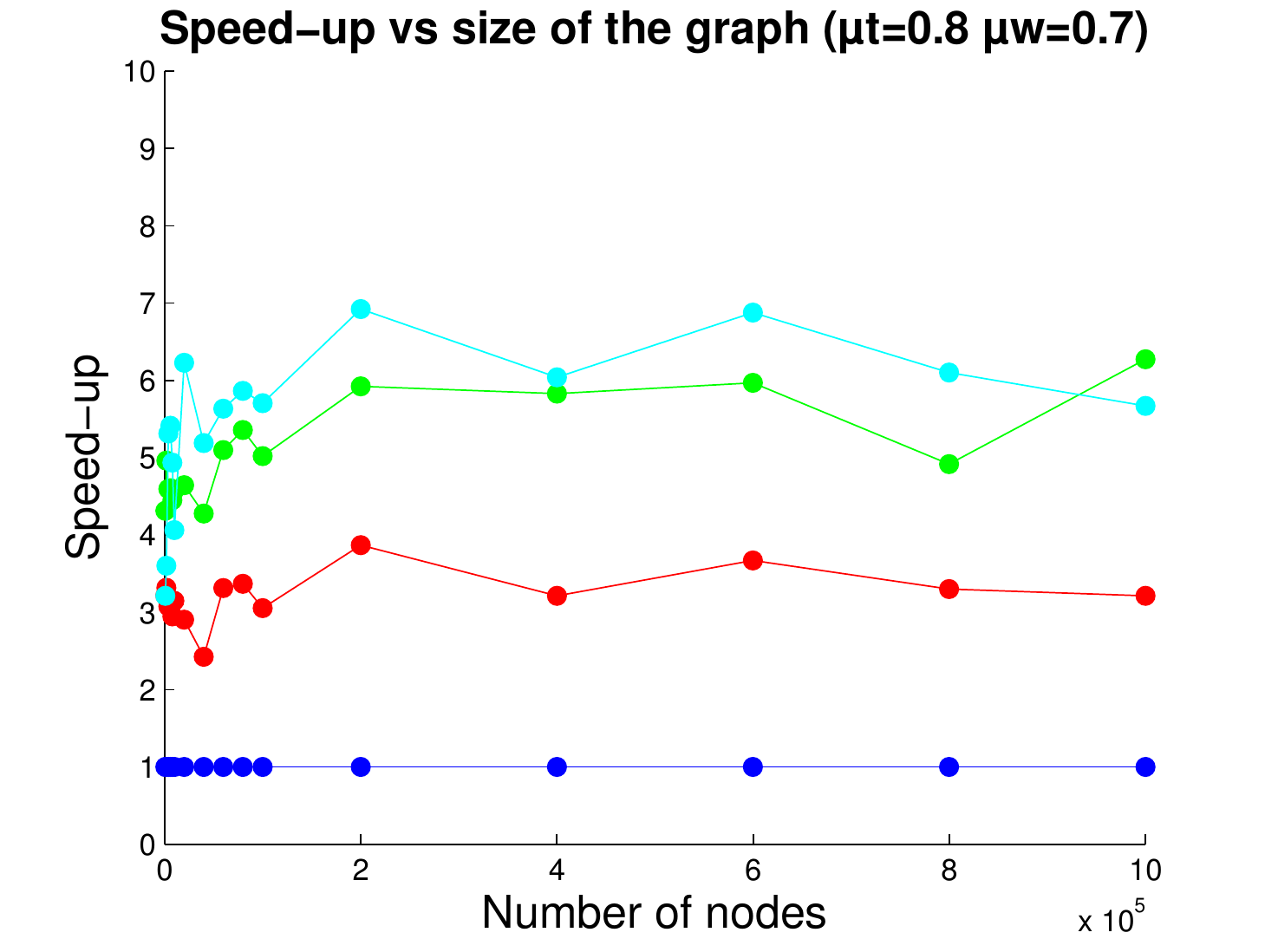}}
    \hspace{5pt}
    \hspace{2.5cm}
  \subfloat{\label{SNd}\includegraphics[scale=1]{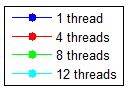}}}
  \caption{Speedup versus number of nodes}
  \label{SN}
\end{figure}

Let us now examine the speedup we actually reach and the influence of the number of threads, the size of the graph and the value of the mixing parameters on it (see Figure \ref{SN}).   \\

The first observation that can be made when looking at Figures \ref{SNa}-\ref{SNc} is that the speedup does not seem to depend on the size of the graph if it is large enough. We have the impression that the speedup asymptotically tends to a constant value (for a constant number of threads), but this phenomenon is not observed for small graphs. Indeed, the computation time for these is smaller so the time "lost" because of overhead is not negligible with respect to the time gained thanks to parallelism. Furthermore, due to a lack of precision in the time measure, any small variation in the computation time can result in an unrealistic great speedup. These two phenomena may explain why the speedup for small graphs seem to be random. \\

Regarding the influence of the number of threads on the speedup, one can clearly see that the speedup increases with the number of threads, for large enough graphs. This is not always the case for smaller graphs, for the same reasons as above.\\

Let us now examine the influence of the mixing parameters on the speedup. It seems that speedup is higher fort greater values of the mixing parameters (at least for large enough graphs). In graphs in which communities are less distinct, more corrections have to be applied before reaching convergence. Given that we parallelised the positive and maximal correction functions, the influence of parallelism is greater than for graphs with lower mixing parameters in which less corrections need to be made.  

To conclude, we can say that the speedup probably depends on the values of the mixing parameters and on the number of threads but not on the size of the graph (for large graphs). It seems that the parallel implementation of Browet's algorithm better outperforms on large graphs with less distinct communities and when the number of threads is higher. We could obtain an average maximum speedup of approximately 7.7 with 12 threads on a graph of $10^6$ nodes and characterised by the mixing parameters $(\mu_t=0.5, \;\mu_w=0.4)$. \\
Given that we used only 2 benchmark graph for each set of parameters, the average value we obtain for the speedup are not very representative. Our results are thus mainly qualitative and tests with more benchmark graphs should be carried out in order to have quantitative results.  


\section{Conclusion}
The aim of this project was to implement a parallel version of Browet's community detection algorithm, in order to get better performances in terms of computation time. First, we had to understand and analyse the different parts of the algorithm to spot the main functions to be parallelised. We determined that three of the four functions constituting the highest hierarchical level could be parallelised.\\

Then, we focused on these functions and their variables. Indeed, to implement the parallel version of Browet's algorithm we chose to use OpenMP. This API requires special attention for the variable status, since it uses shared memory.\\

The results, computed using LFR benchmark graphs, confirm the highly parallelisable structure of Browet's algorithm. Our parallel implementation has an execution time way lower than the sequential one. We also interpreted the role of each parameter involved on the effect of the parallelisation. To conclude, we can clearly consider applying this parallel version of the Synchronised Louvain method on very large networks to highlight their community structure.\\

\paragraph{Future prospects}
The present work sought to give a first approach for the parallelisation of Browet's algorithm. The results are more qualitative than quantitative in the sense that more performance tests should be executed to get more accurate mean computation times. Another interesting work would be to profile the code (for example, with Intel$^{\mbox{\scriptsize{\textregistered}}}$ VTune\texttrademark Amplifier XE) to spot potential parallelisation improvements to be done. We could finally imagine to analyse the classical Louvain method in order to determine if it is parallelisable or not. If so, we could compare performances results of both parallel classical and synchronised Louvain method.

\newpage
\appendix
\section{Appendix}
\subsection{Pseudocodes of main steps of Browet's algorithm}
\label{pseudocodes}
\begin{algorithm}
\caption{ASSIGN}
\label{Assign}
\begin{algorithmic}
\STATE \textbf{function}: ASSIGN(G(V,E))
\FOR{$i \in V$}
\STATE$a(i)=\arg \max_j \Delta H(i\rightarrow j) $
\STATE $\;\;\;\;\;\;\;\;\;\;\;\;\;\;\;\;\;\;\;\;\;\;\rhd$ Best neighbour assignment
\ENDFOR
\STATE $T \leftarrow  (V, \{(i,a(i))\; \forall i\}) \;\;\;\;\;\;\;\;\;\;\;\;\;\;\;\;\;\;\;\;\;\;\;\;\;\;\;\;\;\;\;\;\;\;\;\; \rhd$  Assignment graph
\STATE $C_t \leftarrow WCC(T) \;\;\;\;\;\;\;\;\;\;\;\;\;\;\;\;\;\;\;\;\;\;\;\;\;\;\;\;\;\;\;\;\;\;\;\;\;\;\;\;\;\;\;\;\;\;\rhd$ Weakly Connected Components
\RETURN $C_t$
\end{algorithmic}
\end{algorithm}

\begin{algorithm}
\caption{POSITIVE}
\label{Assign}
\begin{algorithmic}
\STATE \textbf{function}: POSITIVE($C_t$,G(V,E))
\FORALL{$i \in V$}
\STATE $g(i)=-\Delta H(c_i\rightarrow i \rightarrow \{\}) \;\;\;\;\;\;\;\;\;\;\;\;\;\;\;\;\;\;\rhd$ Local gain
\WHILE{$\exists i \in c_i \; \text{with} \; g(i) < 0$}
\STATE $c_1,\; c_2 \leftarrow \; \text{SPLIT}(c_i) \;\;\;\;\;\;\;\;\;\;\;\;\;\;\;\;\;\;\;\;\;\;\;\;\;\rhd$ Optimal assignment graph bisection
\FORALL{$j \in c_1 \cup c_2$}
\STATE $g(j)=-\Delta H(c_j \rightarrow j \rightarrow \{\}) \;\;\;\;\;\;\;\;\;\;\;\rhd$ Update local gain
\ENDFOR 
\STATE $C_t=C_t \setminus \{c_i\} \cup \{c_1,c_2\}\;\;\;\;\;\;\;\;\;\;\;\;\;\;\;\;\;\;\;\;\rhd$ Update communities
\ENDWHILE
\ENDFOR
\RETURN $C_t$
\end{algorithmic}
\end{algorithm}
\begin{algorithm}[H]
\caption{MAXIMAL}
\label{Assign}
\begin{algorithmic}
\STATE \textbf{function}: MAXIMAL($C_t$,G(V,E))
\STATE $C=C_t$
\FORALL{$i \in V$}
\STATE $c_i^*=\arg \max_c \Delta H(c_i \rightarrow i \rightarrow c) \;\;\;\;\;\;\;\;\;\;\;\;\;\;\;\;\;\;\rhd$ Best community for node i 
\ENDFOR
\FORALL{$i \in V,\; \text{if} c_i^* \neq c_i$}
\STATE draw $p(i)$ uniform $\in [0,1]$
\IF {$p(i)<p$}
\STATE$b(i)=branch(i)\;\;\;\;\;\;\;\;\;\;\;\;\;\;\;\;\;\;\;\;\;\;\;\;\;\;\;\;\;\;\;\;\;\;\;\;\;\;\rhd$ Nodes in the tail (branch) of i
\IF{$\Delta H(c_i \rightarrow b(i) \rightarrow c_i^*)>0$}
\STATE $a(i)=\arg \max_{j \in c_i^*}  \Delta H(i\rightarrow j)\;\;\;\;\;\;\;\;\;\;\;\;\;\rhd$ Update assignemnt of i
\STATE $C \leftarrow insert(b(i),c_i^*)\;\;\;\;\;\;\;\;\;\;\;\;\;\;\;\;\;\;\;\;\;\;\;\;\;\;\;\;\;\rhd$ Insert branch $b(i)$
\ENDIF
\ENDIF
\ENDFOR
\RETURN $C$
\end{algorithmic}
\end{algorithm}


\newpage

\renewcommand{\refname}{\spacedlowsmallcaps{References}} 

\bibliographystyle{unsrt}

\bibliography{sample.bib} 


\end{document}